\documentclass[preprint,showpacs,preprintnumbers,amsmath,amssymb]{revtex4}
\usepackage{graphicx}
\usepackage{enumerate}
\usepackage{dcolumn}
\usepackage{bm}
\begin{document}
\title{Properties of Trapped Bose gas with vortices in large-gas-parameter regime }
\author{Arup Banerjee, Manoranjan P. Singh}
\affiliation{Laser Physics Application Section, Centre for Advanced Technology\\
Indore 452013, India}
\begin{abstract}
We study the properties of the vortex state of a trapped Bose gas in the large-gas-parameter regime. To test validity of the Gross-Pitaevskii theory in this regime for the vortex states we compare the results of the Gross-Pitaevskii and the modified Gross-Pitaevskii calculations for the total energy, the chemical potential, the density profile and the frequency shift of the quadrupole modes of the collective oscillations of the condensate. We find that in the large-gas-parameter regime two calculations give substantially different results for all the properties mentioned above. 
\end{abstract}
\maketitle
\section{Introduction}
The mean-field Gross-Pitaevskii (GP) equation has been quite successful in describing both static and dynamic properties of Bose-Einstein condensates (BEC) of alkali atoms confined in magnetic or optical traps \cite{pethik1,pitaevskii1,dalfovo,fetterrv,castinrv}. It is well known that the GP theory is valid for condensates satisfying dilute gas condition $na^{3}<<1$ ( where $n$ is atomic density, $a$ is the s-wave scattering length of interatomic potential and the parameter $x=na^{3}$ is called the gas parameter). Physically it implies that an interacting atomic gas can be considered to be dilute when the average distance between the atoms ($\propto n^{-1/3}$) is much larger than the range of the atomic potential ($\propto a$). Moreover,under this condition the detail shape of the inter-atomic potential is not required rather the boson-boson interaction can be simulated by a pseudo potential which is completely characterised by scattering length $a$. It is important to note here that in some recent theoretical studies  it has been shown that as long as $x\leq 10^{-3}$ the GP theory produces accurate results for both static \cite{nunes,polls1,polls2,banerjee1}and dynamic properties \cite{pitaevskii2,banerjee2}. 
Typically the value of the gas parameters in most of the experiments were in the range $10^{-4}-10^{-5}$ and therefore the GP theory works very well in predicting the properties of these trapped condensates. However, in a recent experiment condensates with peak gas parameter of the order of $10^{-2}$ have been achieved by enhancing the scattering length $a$ with the help of Fesbach resonance \cite{cornish}. For condensates with such large values of the gas parameter the question of validity of GP theory has been raised and tested \cite{polls2,banerjee2} by employing modified GP (MGP) theory within the local-density approximation. These studies clearly show that for large values of gas parameter ($\approx 10^{-2}$) both ground-state properties ( energy, chemical potential) and dynamic properties ( frequencies of collective oscillations) are significantly modified by MGP theory.

The vortex states in trapped condensates play an important role in establishing the superfluid properties of the condensates \cite{pethik1,pitaevskii1}. It is then natural to ask how the properties of vortex states are modified in the large-gas-parameter regime. This has motivated us to study the properties of the trapped condensates with vortex states in the large-gas-parameter regime. To describe the condensate in the large-gas-parameter regime we employ the MGP theory which is considered to be applicable in this regime. By comparing the MGP results with the corresponding GP numbers we make a systematic assessment of the corrections introduced by the MGP equation over the GP results. In this paper we consider condensates with the values of gas parameter spanning a range similar to that achieved in the experiment of Cornish et al. \cite{cornish}. The results presented in this paper can be divided in to two parts. In the first part 
we focus our attention on the modifications of energy, chemical potential and the density profile of the condensate with a quantized vortex. 

The quantized vortex states not only affect the static properties of the condensate but the dynamic properties also get modified by them. For example, the  presence of a quantized vortex state leads to the splitting of the frequencies of the two modes of quadrupole 
oscillations with opposite values of the third component of the 
angular momentum, which are degenerate in the absence of a quantized 
vortex state \cite{zambelli,svidzinsky,graham,haljan}.  
The vortex state in a condensate breaks the time reversal 
symmetry which in turn results in the removal of the degeneracy 
of the two modes of oscillations carrying opposite values of angular 
momentum. Therefore, the splitting of the two quadrupole modes of 
collective oscillations can be employed to detect the presence of 
a quantized vortex states in the BEC \cite{haljan} as the measurement of the frequencies of the collective oscillations can be carried out with high precision . 
In the second part of this paper we focus our attention on the 
effect of the large gas parameter on the splitting of two 
degenerate quadrupole modes due to the presence of a vortex state. We calculate and the compare the results for the frequency shift of the two quadrupole modes by employing the GP and the MGP equations in the large-gas-parameter regime. 

For the purpose of calculation we use the variational approach with quite accurate variational ansatz for the wave function of a trapped condensate with a quantized vortex line along the z-axis \cite{mpsingh} and the frequency shifts of the quadrupole modes are then obtained by employing well established sum-rule approach \cite{stringari,bohigas}.

The paper is organized in the following manner. In section II we describe the theoretical methods employed in this paper and briefly describe the MGP theory followed by the variational method employed to obtain the wave function of the vortex state and other physical observables mentioned above and the sum-rule approach for the calculation of frequencies of collective oscillation. The section III is devoted to the discussion of the results. The paper is concluded in section IV.
  
\section{Theory}
The ground-state energy functional associated with a condensate of N bosons each with mass $m$ confined in a trap potential $V_{t}({\bf r})$  can be written as \cite{nunes}
\begin{equation}
E[\Psi] = \int d{\bf r}\left [ \frac{\hbar^{2}}{2m}|{\bf \nabla}\Psi|^{2} +
V_{t}({\bf r})|\Psi|^{2} + \epsilon(n)|\Psi|^{2}\right ].
\label{1}
\end{equation}
where $\Psi({\bf r})$ is the the condensate wave function (order parameter) and $n({\bf r})$ represents the corresponding density and it is given by $n({\bf r}) = |\Psi({\bf r})|^{2}$ . The condensate wave function $\Psi ({\bf r})$ can be determined by minimizing the above energy functional. In the above equation the first, the second and the third terms  represent the kinetic energy of bosons, the energy due to the trapping potential, and the inter-atomic interaction energy within the local density approximation (LDA), respectively. To go beyond the GP theory we make use of the perturbative expansion for $\epsilon(n)$
in terms of the gas parameter $na^{3}$,
\begin{equation}
\epsilon(n) = \frac{2\pi\hbar^{2}an}{m}\left [1 + \frac{128}{15\sqrt{\pi}}\left (na^{3}
\right )^{\frac{1}{2}} + 8\left (\frac{4\pi}{3} - \sqrt{3}\right )(na^{3})\ln\left (na^{3}\right )
+ {\cal O}\left(na^{3}\right)\right ].
\label{2}
\end{equation}
The first term in the above expansion, which corresponds to the
energy of the homogenous Bose gas within the mean-field theory as
considered in the GP theory, was calculated by Bogoliubov
\cite{bogoliubov}. The second term was obtained by Lee, Huang and
Yang (LHY) \cite{lee}, while the third term was first calculated
by Wu \cite{wu} using the hard-sphere model for the interatomic
potential.  Although, it has been emphasized in the literature that the above expansion
is valid only for $na^{3}<<1$, it is only recently that the range of validity of above expansion has been systematically
investigated by Giorgini et al. \cite{giorgini}. They have used diffusion
Monte Carlo (DMC) method to calculate the ground-state of uniform
gas of bosons interacting through different model potentials with
$a>0$. It has been found that the expansion (\ref{2}) is valid as
long as $na^{3}<10^{-3}$. However, for gas parameter beyond
$10^{-3}$ the inclusion of the logarithmic term leads to a severe
mismatch with the DMC simulation results. On the other hand, expansion (\ref{2})
up to the LHY term gives an accurate representation of the DMC calculations even for the gas
parameter of the order of $10^{-2}$.  Consequently, we do not
consider the logarithmic term in the expansion (\ref{2}) for all the calculations in this study.

The trapping potential $V_{t}({\bf r})$ is taken to be
axially symmetric characterized by two angular frequencies $\omega_{\bot}^{0}$ and
$\omega_{z}^{0}$ ($\omega_{x}^{0}=\omega_{y}^{0}=\omega_{\bot}^{0}\neq\omega_{z}^{0}$). It is given by
\begin{equation}
V_{t}({\bf r}) = \frac{m{\omega_{\bot}^{0}}^{2}}{2}\left (x^{2} +
y^{2} + \lambda_{0}^{2}z^{2} \right ), \label{3}
\end{equation}
where $\lambda_{0} = \omega_{z}^{0}/\omega_{\bot}^{0}$ is the anisotropy parameter of
the trapping potential ($\lambda_{0}=1$ corresponds to a spherically symmetric trap).

The ground-state energy functional (\ref{1}) can be easily generalized to include the vortex states. In this paper we consider condensates with a vortex line along the z-axis and all the atoms are flowing around it. The wave function for such a condensate can be written as \cite{dalfovo1} 
\begin{equation}
\Psi({\bf r}) = \psi({\bf r})e^{i\kappa\phi}
\label{vortexwavefunction}
\end{equation}
where $\phi$ is the angle around the z-axis and $\kappa$ is an integer denoting the quantum circulation and the total angular momentum along the z-axis is given by $N\hbar\kappa$. Substituting the above complex wave function of Eq.(\ref{vortexwavefunction}) in Eq. (\ref{1}) we get generalized energy functional for the condensate with a vortex state as \cite{dalfovo1}
\begin{equation}
E[\psi ] = \int d{\bf r}\left
[\frac{\hbar^{2}}{2m}|{\bf\nabla}\psi|^{2} + \frac{\hbar^{2}\kappa^{2}}{2mr_{\bot}^{2}}|\psi|^{2} +
v_{ext}({\bf r})|\psi|^{2} + \epsilon (\rho )|\psi|^{2}\right ].
\label{energyfuncvortex}
\end{equation} 
In the above equation $r_{\bot} = \sqrt{x^{2} + y^{2}}$ and the density $n({\bf r}) = |\psi({\bf r})|^{2}$. The the presence of centrifugal term due to the vortex state makes the above functional different from Eq. (\ref{1}).  The minimization of the above functional with respect to $\psi(r)$ with the constraint 
\begin{equation}
\int|\psi ({\bf r})|^{2}d{\bf r} = N
\label{norm}
\end{equation}
leads to the MGP equation for the condensate with vortex state
\begin{equation}
 \left [- \frac{\hbar^{2}}{2m}{\bf \nabla}^{2} + \frac{\hbar^{2}\kappa^{2}}{2mr_{\bot}^{2}} +
V_{t}({\bf r}) + \frac{4\pi\hbar^{2}a}{m}|\psi|^{2}\left (1 + \frac{32a^{3/2}}{3\pi^{1/2}}|\psi|^{2}\right )\right ]\psi({\bf r}) = \mu\psi({\bf r})
\label{mgpeq}
\end{equation}
where $\mu$ is the chemical potential arising from the constraint condition given by Eq. (\ref{norm}). The GP equation for the vortex state can be obtained from Eq. (\ref{mgpeq}) by neglecting the interaction energy term proportional to $|\psi|^{4}$ in this equation. It is not possible to find an exact solution to Eq. (\ref{mgpeq}). Consequently various numerical techniques have been developed to solve the above nonlinear Schrodinger equation for studying the properties of trapped BEC \cite{bao,minguzzi}. In this paper we use the variational method to obtain the wave function (or density) and other relevant physical observables of the condensate. The main advantage of this method is that with a suitable choice for the variational form for the wave function one can get quite accurate results with less computational effort. In our earlier work \cite{mpsingh,banerjee1,banerjee2} we have demonstrated the applicability of the variational method by calculating the ground state properties and the collective oscillation frequencies of condensates both with the GP and the MGP equations for wide range of values of  particle number and scattering length. 

Following Ref. \cite{mpsingh} the variational form for the wave function of the condensate with quantized vortex state is chosen as
\begin{equation}
\psi({\bf r}_{1}) =  Ar_{1 \bot}^{q}e^{-\frac{1}{2}\left(\frac{\omega_{\bot}}{\omega_{\bot}^0}\right)^p
\left( r_{1 \bot}^2 +\lambda z_1^2\right)^p},
\label{varden}
\end{equation} 
where $q$, $\lambda$, $\omega_{\bot}$ and $p$ are the variational
parameters which are obtained by minimizing the energy functional given by \ref{energyfuncvortex} with respect to
these parameters. In the above equation we use the scaled
co-ordinate ${\bf r}_{1}={\bf r}/a_{ho}$ with $a_{ho} = \left
(\hbar/m\omega_{\bot}^{0}\right )^{\frac{1}{2}}$. We note that
$n({\bf r_{1})}$ is normalized to unity and from Eq. (\ref{norm})
it can be seen that
\begin{equation}
n({\bf r}) = \frac{N}{a_{h0}^{3}}n({\bf r_{1})}.
\end{equation}
The above variational form of the wave function  has already been shown to describe  the  state of the dilute Bose gas with a quantized vortex  confined in a trap quite accurately for a wide range of particle numbers \cite{mpsingh}. In this paper we extend the applicability of this wave function in the large-gas-parameter regime also.

The proportionality factor $A$ in Eq. (\ref{varden}) is determined by the normalization condition (Eq. (\ref{norm}))
\begin{equation}
A^{2} = \frac{\sqrt{\lambda}\Gamma (3/2 + q)}{\pi^{3/2}\Gamma(1 + q)\Gamma((3/2 +q)/2p)}\left (\frac{\omega_{\bot}}{\omega_{\bot}^0}\right)^{(3/2 + q)}
\end{equation}
Using this variational form for the wave function physical observables can be expressed analytically in terms of the four variational parameters. For example, the energy functional
$E_{1} = E/\hbar\omega_{\bot}^{o}$ can be written as 
\begin{equation}
\frac{E_{1}}{N} = T + U + E_{rot} + E^{1}_{int} + E^{2}_{int}.
\label{5}
\end{equation}
Here $T$, $U$ and $E_{rot}$ denote the average kinetic, the trapping
potential and the rotational energies per particle, respectively. The term
E$^{1}_{int}$ gives the interaction energy per particle in the
mean-field approximation as considered in the GP theory while
E$^{2}_{int}$ gives the correction due to the LHY term in the
expansion (\ref{2}) resulting in the MGP equation. The analytical expressions for these energy components are
\begin{equation}
T = \frac{\omega_{\bot}(1 + 2q)\left [(1 + 2p)(1 + \lambda/2) + q(2p + 2q +\lambda)\right ]\Gamma((1 + 2q)/2p)}{\omega_{\bot}^{0}4(3 + 2q)\Gamma((3 + 2q)/2p)},
\label{kinenergy}
\end{equation}
\begin{equation}
U = \frac{\omega_{\bot}^{0}\left (1 + q + \lambda_{0}^{2}/2\lambda\right )\Gamma((5 + 2q)/2p)}{\omega_{\bot}(3 + 2q)\Gamma((3 + 2q)/2p)},
\label{tarpenergy}
\end{equation}
\begin{equation}
E_{rot} = \frac{\kappa^{2}\omega_{\bot}(1 + 2q)\Gamma((1 + 2q)/2p)}{4\omega_{\bot}^{0}q\Gamma((3 + 2q)/2p)}
\label{rotenergy}
\end{equation}
\begin{equation}
E^{1}_{int} = 2N\tilde{a}\left( \frac{\omega_{\bot}}{\omega_{\bot}^{0}}\right)^{3/2}\frac{p\sqrt{\lambda}\Gamma^{2}(q + 3/2)\Gamma(2q + 1)\Gamma((4q + 3)/2p)}{2^{(4q + 3)/2p}\sqrt{\pi}\Gamma^{2}((2q + 3)/2p)\Gamma^{2}(q + 1)\Gamma(2q + 3/2)},
\label{intenergy1}
\end{equation}
and 

\begin{eqnarray}
E^{2}_{int} & = &  \frac{128}{15\sqrt{2}\pi}\frac{2\sqrt{2}}{\pi^{9/4}}N^{3/2}\tilde{a}^{5/2}\left( \frac{\omega_{\bot}}{\omega_{\bot}^{0}}\right)^{9/4}\left (\frac{2}{5}\right)^{(5q + 3/2p)}p^{3/2}\lambda^{3/4}\nonumber \\
& \times & \frac{\Gamma^{5/2}(q + 3/2)\Gamma(5q/2 + 1)\Gamma((5q + 3)/2p)}{\Gamma^{5/2}((2q + 3)/2p)\Gamma^{5/2}(q + 1)\Gamma((5q + 3)/2)}
\label{intenergy2}
\end{eqnarray} 
where $\tilde{a} = a/a_{ho}$ and $\Gamma(m)$ is the gamma function. It is easy to check that above energy components correctly reduce to the expressions derived in \cite{banerjee1} for the case of condensate without quantized vortex by setting $\kappa$ and $q$ equal to zero. For particular values of $N$ and $\tilde{a}$ the parameters $\omega_{\bot}$, $q$, $\lambda$ and $p$ are determined by minimizing the energies given by Eq. (\ref{5}) and Eqs. (\ref{kinenergy})-(\ref{intenergy2}).
 
Having described the variational method for obtaining the wave function and the other physical properties we now describe the sum-rule approach of 
many-body response theory for the calculation of frequencies of 
collective oscillations. In the collisionless regime the collective 
excitation frequencies of a confined bosonic gas is well described 
by the sum rule approach. The most important advantage of this 
approach is that the calculation of frequencies requires the 
knowledge of the ground-state wave function (or the ground-state 
density) of the many-body system only. This method has been 
extensively applied to calculate the frequencies of the collective 
oscillations of trapped atomic gases \cite{pitaevskii1}. In this 
paper we follow the sum-rule approach developed in \cite{zambelli} 
for the calculation of the splitting of the quadrupole modes of 
collective oscillations. 

The collective excitations of any 
many-body system are generally probed by applying an external 
excitation field. For excitation operators $F_{+}$ and $F_{-}$ 
exciting two quadrupole modes with opposite angular momentum, 
respectively, the corresponding strength functions are given 
by \cite{bohigas,lipparini}
\begin{equation}
S_{\pm}(E) = \sum_{n}|\langle 0|F_{\pm}|n\rangle |^{2}\delta(E- E_{n})
\label{strength}
\end{equation}
where $|0\rangle$ denotes the ground state of the system and 
$E_{n}-E_{0}$ is the excitation energy of the eigenstate 
$|n\rangle$ of the Hamiltonian $H$ relative to the ground state 
energy $E_{0}$. The $p$-th order moments of the strength function 
can be defined as
\begin{equation}
m_{p}^{\pm} = \int E^{p}\left ( S_{+}(E) - S_{-}(E)\right )dE .
\label{pthmoment}
\end{equation}
These moments provide various energy weighted sum rules which are 
employed to obtain the frequencies of collective oscillations. 
One of the important properties of these moments is that, for 
a given $p$, some of the moments  can be expressed in terms of 
the commutators of the excitation operators with the many-body 
Hamiltonian $H$. Some of the energy weighted sum rules which 
are relevant for this paper are
\begin{eqnarray}
m_{0}^{-} & = & \langle 0|\left[F^{\dagger}, F\right ]| 0\rangle \nonumber \\
m_{1}^{+} & = & \langle 0|\left [F^{\dagger},\left [H, F\right ]\right ]|0\rangle \nonumber \\
m_{2}^{-} & = & \langle 0|\left [ \left [F^{\dagger}, H\right ],\left [H, F\right ]\right ]|0\rangle \nonumber \\ 
m_{3}^{+} & = & \langle 0|\left [ \left [F^{\dagger}, H\right ],\left [H,\left [H, F\right ]\right ]\right ]|0\rangle
\label{moments1}
\end{eqnarray}
where $|0\rangle$ represents the ground state of the hamiltonian $H$ and $[,]$ represents the commutator between the corresponding 
operators. 

These expressions can be used to calculate the frequency shifts of 
the quadrupole oscillations of a trapped condensate. For the quadrupolar case we will consider the modes excited by the operators
\begin{equation}
F_{\pm}  = (x \pm iy)^{2}
\label{excitationop} 
\end{equation}
carrying angular momentum $m_{z} = \pm 2$, where $m_{z}$ represents the z-component of the angular momentum of the elementary excitation. 
In order to use the above expressions for the calculation of 
the shift in frequencies of quadrupole modes we now assume 
that the moments $m_{p}^{\pm}$ are exhausted by a single 
excitation with frequency $\omega_{\pm}$ and strength 
$\sigma_{\pm}$. Under such single-mode approximation the 
strength distribution can be written as
\begin{equation}
S_{\pm}(E) = \sigma_{\pm}\delta (E - \hbar\omega_{\pm}).
\label{singlemode}
\end{equation}
Moreover, the vanishing of $m_{0}^{-}$ ( as given by the first moment of Eq. (\ref{moments1})) due to the commutator of $F_{+}$ and $F_{-}$ being zero leads to the result $\sigma_{+} = \sigma_{-} = \sigma$. We note here that the single 
mode approximation is well suited for the condensates with 
large number of atoms $N$ and positive scattering length 
$a$ \cite{zambelli}. Using Eqs.\,(\ref{singlemode}) 
and (\ref{pthmoment}) we obtain the following expression for the 
difference between the two frequencies  
\begin{equation}
\hbar\delta = \hbar\left (\omega_{+} - \omega_{-}\right ) = \frac{m_{2}^{-}}{m_{1}^{+}}.
\label{difference}
\end{equation}  
Furthermore, the individual frequencies of the quadrupole 
oscillations can be determined by using Eq. (\ref{difference}) along with the expression for the square of the mean frequency given by
\begin{equation}
\hbar^{2}\left (\frac{\omega_{+} + \omega_{-}}{2}\right )^{2} = 
\frac{m_{3}^{+}}{m_{1}^{+}} - 
\frac{3}{4}\left (\hbar\delta\right)^{2}\, .
\label{mean}
\end{equation}
In the next section we calculate the moments $m_{1}^{+}$, $m_{2}^{-}$ and $m_{3}^{+}$ in terms of some averages over the vortex state wave function and use Eqs. (\ref{difference}) 
and ({\ref{mean}) to determine the frequency shifts
of the two quadrupole modes due to the presence of a quantized vortex in a trapped condensate falling in the large gas parameter regime.
\section{Results and Discussion}
\subsection{Ground state properties}
We begin this section with a comparison of our results obtained by the variational method with the corresponding numbers of Ref. \cite{nilsen} calculated by solving the MGP equation by steepest descent methods. We make this comparison to check the accuracy of the variational approach as it has been shown in Ref. \cite{nilsen} that their MGP results are in very good agreement with the results of ab-initio variational Monte Carlo calculation both in the presence and in the absence of vortex state. At this point we note that in addition to the comparison with the ab-initio results the virial relation among the different energy components also provides a way of checking the correctness and the accuracy of the variational solutions. For the MGP theory in the presence of a single quantized vortex the virial relation is given by
\begin{equation}
2T - 2U + 2E_{rot} + 3E^{1}_{int}  +\frac{9}{2}E^{2}_{int} = 0
\label{virial}
\end{equation}
This expression has been derived by using the variational nature of the ground state energy and the scaling transformation $n({\bf r_{\eta}})\longrightarrow\eta^{3}n({\bf r})$. In all our calculations the virial relation given above is satisfied up to sixth decimal place or better. This indicates that the the variational method employed in this paper yields quite accurate results for condensates with vortex state in the large-gas-parameter regime. 

For the purpose of comparison we calculate the total energy, individual energy components given by Eqs. (\ref{kinenergy})- \ref{intenergy2}) and the chemical potential of a $^{87}Rb$ condensate carrying a single vortex state ($\kappa = 1$). In accordance with the Ref. \cite{nilsen} we consider a disk-shaped trap with $\lambda_{0} = \sqrt{8}$ and $\omega_{z}^{0} = 2\pi\times 220$ Hz confining $N = 500$ condensate atoms and the scattering length is chosen to be $a = 0.15155a_{ho}$.  The results of this calculation are shown in Table I along with the corresponding numbers of Ref. \cite{nilsen} in parentheses. 
It can be clearly seen from Table I that numbers obtained by the variational method are very close to the corresponding results of Ref. \cite{nilsen}. In particular our numbers for the total energy, trap energy and both the components of the interaction energies are slightly higher (of the order of $1\%$) than those of Ref. \cite{nilsen}. On the other hand, our results for the kineteic energy and the rotaional energy are appreciably lower and higer respectively, than the corresponding energies reported in Ref. \cite{nilsen}. Despite the  difference between the results for the kinetic and the rotational energies the results obtained for the total energy and the chemical potential obtained by the two metods match quite well due to cancellation of errors. The cancellation of errors also lead to very good satisfaction of the virial relation. Moreover, we find that for the above choice of $N$ and $a$ the MGP reuslts for the total energy and the chemical potential are approximately  $19\%$ and $16\%$ respectively, higher than the corresponding the GP numbers.  These results along with our earlier work \cite{mpsingh,banerjee1,banerjee2} clearly demonstrate the applicability of the variational approach in calculating the properties of condensates both with and without quantized vortex state in the large gas parameter regime. 

Now we focus our attention on the application of the MGP theory to condensates with a single quantized vortex for larger values of gas parameter as achieved in the experiment of Cornish et al. \cite{cornish}. In accordance with the experiment we consider $N = 10^{4}$ $^{85}Rb$ atoms confined in an anistropic trap with frequencies $\omega_{\bot}^{0} = 2\pi\times17.5$ Hz and $\omega_{z}^{0} = 2\pi\times 6.9$ Hz and the scattering length is varied from $a = 1400a_{0}$ to $a = 10000a_{0}$, where $a_{0}$ is the Bohr radius of hydrogen atom. We note here that this range of values of the scattering length falls in the large gas parameter regime as the maximum value of $a$ corresponds to the peak gas parameter $x_{peak}\approx 10^{-2}$. In Table II we show the results for the total energy and the chemical potential obtained by employing the MGP equation and compared them with the corresponding numbers calculated by the GP equation. It clearly shows that the difference between the MGP and the GP results increases with increase in the scattering length. For $a/a_{0} = 1400$ the differences in the total energy and the chemical potential are approximately $3\%$ and $4\%$ respectively.  On the other hand, for the maximum value of $a/a_{0} = 10,000$ the MGP reults for the total energy and the chemical potential are higher by around $26\%$ and $30\%$ respectively, over the corresponding GP numbers. From these results we conclude that in the large gas parameter regime the MGP results for the condensates with a single quantized vortex are substantially altered as compared to the corresponding GP results. It is important to note here that although the MGP corrections to the energy are significant, the critical angular velocity $\Omega_{c}$ for the formation of the vortex state given by \cite{dalfovo1}
\begin{equation}
\frac{\Omega_{c}}{\omega_{\bot}^{0}} = \kappa^{-1}\left [\left (E_{1}/N\right )_{\kappa} -  \left (E_{1}/N\right )_{0}\right ],
\end{equation}
changes nominally as compared to the GP calculations. For example, corresponding to $a/a_{0} = 10,000$ we obtain $\Omega_{c} = 0.325\omega_{\omega_{\bot}}$ and 
$\Omega_{c} = 0.335\omega_{\omega_{\bot}}$ for a single vortex with $\kappa = 1$ by employing the MGP and the GP calculations respectively. This observation regarding the small change in value of critical angular velocity was also made in Ref. \cite{nilsen}, however, for much lower value of the gas parameter. Our results clearly shows that the MGP calculation does not provide any sizable correction to the critical angular velocity $\Omega_{c}$ even for very large value of the gas parameter.

Having estimated the corrections introduced by the MGP calculation over the GP results for the energy and the chemical potential now we wish to present the results for the density profile obtained by the MGP and the GP calculations and compare them to study the effect of large gas parameter on the density profile. In Fig. 1 and 2 we show the density profile of the condensates with a single vortex for the values of scattering length $a/a_{0} = 1400$ and $a/a_{0} = 10000$ respectively obtained by the  MGP and the GP calculations. It can be clearly seen from these two figures that for $a/a_{0} = 1400$ the MGP calculation leads to slight modification in the density profile over the GP result. On the other hand, for $a/a_{0} = 10000$ the MGP density profile is substantially different from the corresponding GP density profile. Due to the repulsive nature of the LHY term in the MGP equation the bosons in the condensates are pushed outward resulting in the increase in the size of the condensate and decrease in the value of maximum density in comparison to the GP case. Furthermore, it is evident from Fig.2 that the values of the peak density $n_{peak}$ and the the transverse distance $r_{1\bot}^{peak}$ at which densities become maximum are significantly different in the GP and the MGP calculations .  To show the difference in the MGP and the GP density profiles in more quantitative manner we plot in Fig. 3 and 4 the peak density $n_{peak}$ and the corresponding peak position $r_{1\bot}^{peak}$ respectively, as functions of the scattering length. From the variational form of the wave function given by Eq. (\ref{vortexwavefunction}) it is easy to verify that the density peaks at
\begin{equation}
r_{1\bot}^{peak} = \sqrt{\frac{\omega_{\bot}^{0}}{\omega_{\bot}}}\left (\frac{q}{p}\right )^{1/2p}
\label{rpeak}
\end{equation}
and the expression for the peak density is given by
\begin{equation}
n_{peak} = A^{2}\left (\frac{\omega_{\bot}^{0}}{\omega_{\bot}}\right )^{q}\left (\frac{q}{p}\right )^{q/p}e^{-q/p}
\label{peakdensity}
\end{equation}
The values of variational parameters $\omega_{\bot}$, $\lambda$, $p$ and $q$ obtained by minimization of the energy are used to calculate  $r_{1\bot}^{peak}$ and $n_{peak}$ by using Eqs. (\ref{rpeak}) and (\ref{peakdensity}). From  Fig. 3 we observe that with  both  MGP and the GP calculations the peak density shows a decreasing trend with increasing $a/a_{0}$ due to greater repulsion between the bosons. As mentioned above for the MGP calculation the second term in inter-atomic potential corresponds to more repulsion which results in higher reduction in the peak density as compared to the GP case. The difference between the MGP and the GP peak densities increases with increasing scattering length. For example, as $a/a_{0}$ is scanned from $1400$ to $10000$ the difference between the GP and the MGP results grows from around $7\%$ to $37\%$. 
In contrast to this $r_{1\bot}^{peak}$ increases as the scattering length is increased in conformity with the fact that the size of the condensate grows as the repulsive interaction goes up. However, like peak density the difference between the results for $r_{1\bot}^{peak}$ by the MGP and the GP calculations also increases with increasing value of $a/a_{0}$. For $a/a_{0} = 1400$ and $a/a_{0} = 10000$ the differences in the MGP and the GP $r_{1\bot}^{peak}$ are found to be approximately $3\%$ and $22\%$ respectively. Furthermore we also notice from Fig. 1 and 2 that the depth of the vortex hole around the z-axis obtained by the GP calculation is significantly greater than the MGP case in the large-gas-parameter regime.  However, the results for the size of the hole around z-axis  characterised by the ratio $r_{1\bot}^{peak}/\sqrt{\langle 0|x^{2} + y^{2}|0 \rangle}$ \cite{dalfovoexpansion}, obtained by the two calculations show difference of lesser degree as the scattering length is increased. For example corresponding to $a/a_{0}=1400$ and $a/a_{0}=10000$ the differences in the size of the holes are of the order of $1\%$ and $4.5\%$ respectively.

Thus we conclude that in presence of vortex the MGP results for the  properties like the total energy, chemical potential and the total density profile except the critical angular velocity $\Omega_{c}$ and the size of the vortex hole around the z-axis  differ significantly in comparison to the GP predictions in the large-gas-parameter regime.
In the next section we discuss the results for the splitting in the quadrupole modes of collective oscillations due to the presence of vortex in the large gas parameter regime.
\subsection{Splitting of quadrupole modes}
To calculate the frequency shifts of the two quadrupole modes characterized by the components of angular momentum $m_{z} = 2$ and $m_{z} = -2$ respectively, first we need to evaluate the moments $m_{1}^{+}$, $m_{2}^{-}$ and $m_{3}^{+}$.
For excitation operators given by Eq. (\ref{excitationop}) the commutators in  Eq. (\ref{moments1}) 
can be evaluated to give following expressions for the moments \cite{zambelli,castin}
\begin{eqnarray}
m_{1}^{+} & = & \frac{8\hbar^{2}}{m}\langle 0|x^{2} + y^{2}|0\rangle \nonumber \\
m_{2}^{-} & = &  \frac{16\hbar^{3}}{m^{2}}\langle 0|xp_{y} - yp_{x}|0\rangle  \nonumber \\ 
m_{3}^{+} & = & \frac{16\hbar^{4}(\omega_{\bot}^{0})^{2}}{m}\left [\langle 0|x^{2} + y^{2}|0\rangle + \frac{\langle 0|p_{x}^{2} + p_{y}^{2}|0\rangle}{m^{2}(\omega_{\bot}^{0})^{2}} \right ]
\label{moments}
\end{eqnarray}
where $p_{i}$ is the $i$-th component of the linear momentum 
vector ${\bf p}$ and $|0\rangle$ represents the wave function of the vortex state obtained by solving the time independent MGP or GP equations (Eq. (\ref{mgpeq})). In deriving above equations the relation 
$F_{+}^{\dagger} = F_{-}$ has been used. It is worth noticing 
that the moments up to third-order do not depend on the boson-boson 
coupling parameter (two-body interaction) explicitly. This is 
because the energies within the LDA do not contribute to moments 
up to third order for the excitation operators satisfying 
${\bf\nabla^{2}}F_{\pm} = 0$ \cite{liserra}. 
Now use of moments $m_{1}^{+}$ and $m_{2}^{-}$ in Eqs.  (\ref{moments}) and (\ref{difference})yields the result
\begin{equation}
\delta = \frac{2}{m}\frac{N\hbar\kappa}{\langle r_{\bot}^{2}\rangle}
\label{splitting}
\end{equation}
where $\langle r_{\bot}^{2}\rangle$ is the transverse size of the condensate determined by the average
\begin{equation}
\langle r_{\bot}^{2}\rangle = \int \left (x^{2} + y^{2}\right )n({\bf r})d{\bf r}
\end{equation}
It is important to note that albeit the frequency shift does not 
explicitly depend on the boson-boson coupling parameter but 
implicit dependence on the two-body interaction enters through 
the wave function or the density of the condensate 
which crucially depend on the nature of the boson-boson interaction.
We have already observed  that the transverse sizes of the condensate  obtained by employing the MGP and the GP equations differ significantly in the large-gas-parameter regime. Consequently, the MGP and the GP frequency shifts of the two quadrupole modes will also vary significantly in this regime. 
We show the result for the MGP and the GP frequency shifts as a function of the scattering length in Fig. 5. The frequency shift between the two quadrupole modes decreases with the increase in the scattering length due to the fact that higher values of scattering length correspond to greater repulsive interaction between the bosons resulting in condensates with larger values of $\langle r_{\bot}^{2}\rangle$. For up to around $a/a_{0}=1000$ the GP and the MGP frequency shifts are nearly identical, however, as the scattering length increases further the difference between the two results also increases. For example, at the maximum value of the scattering length ($a/a_{0} =10000$) we find that the GP frequency shift is around $26\%$ higher in comparison to the MGP result. 

Finally, we  note that the quadrupole frequencies $\omega_{\pm}$ can be obtained from Eq. (\ref{mean}) and the expressions for the $m_{1}^{+}$ and $m_{3}^{+}$ given in Eq. (\ref{moments}). For the values of the gas parameter considered in this paper the transverse kinetic energy $\langle 0|p_{x}^{2} + p_{y}^{2}|0\rangle$ is much much smaller than the transverse radius $\langle 0|x^{2} + y^{2}|0\rangle$ of the condensate with a vortex state  and thus it can be neglected in $m_{3}^{+}$. As a result of this, under single mode approximation (Eq. (\ref{singlemode})) the expressions for the quadrupole frequencies are given by
\begin{equation}
\omega_{\pm} = \sqrt{2}\omega_{\bot}^{0} \pm \frac{\delta}{2}
\label{quadrupolefreq}
\end{equation}

From the above results we conclude that in the large-gas-parameter regime the MGP and the GP results for the frequency shift of the two quadrupole modes due to the presence of a quantized vortex and also the frequencies of the two modes are substantially different for the values of scattering lengths which are achieved by tuning the Fesbach resonance.
\section{Conclusion}
In this paper we have studied the properties of BEC with single quantized vortex in the large gas parameter regime. For this purpose we have employed MGP equation which has been obtained by including the second term (LHY term) in the perturbative expansion of the inter-atomic interaction energy per particle obtained from the ground-state energy of uniform Bose gas. We have used the variational approach to solve the MGP equation by employing a suitable ansatz for the wave function of a condensate with a quantized vortex.
The wave function obtained by variational approach is then employed to calculate the static properties like the total energy, chemical potential and the density profile for wide range of scattering length lying well within the large gas parameter regime. The correctness and accuracies of our solutions are checked by  verifying the generalized virial relation and also comparing them with the solutions obtained by the steepest descent method. By using the wave function of the vortex state and the sum-rule approach we have also calculated the frequency shift of the two quadrupole modes of the collective oscillations due to the presence of a quantized vortex in the large gas parameter regime. To test the accuracy of GP equation we have made a detailed comparison of the results of the GP and the MGP equations for all the observables mentioned above. We have found that in the large gas parameter regime MGP equation provide substantial amount of corrections in all the observables considered in this paper. These changes are quite sufficient to be observed in measurements 
involving condensates with large gas parameter as achieved in
Ref. \cite{cornish}. The comparison of the results obtained in this paper with the experimental results will also test the validity of the GP theory for the description of such condensates.
\section{Acknowledgment} We wish to thank Dr. S. C. Mehendale for critical reading of the manuscripts and a helpful discussions.
 
\newpage
\begin{table}
\caption{Results for the chemical potential and the energies in units of $\hbar\omega_{\bot}^{0}$ obtained from the GP and the MGP calculations for the condensate of of $N = 500$ $^{85}$Rb atoms trapped in an anisotropic trap
with $\lambda_{0} = \sqrt(8)$ and $\omega_{z}^{0}= 2\pi\times 220Hz$. The scattering length is $\tilde{a} = a/a_{\bot} = 0.15155$. Numbers in parenthesis are results
of Ref. \cite{nilsen}}
\tabcolsep=0.15in
\begin{center}
\begin{tabular}{|c|c|c|c|c|c|c|c|}
\hline
$ $ & $\mu_{1}$ & $E_{1}/N$ & $T/N$ & $U/N$ & $E_{int}^{1}/N$ & $E_{int}^{2}/N$ &  $E_{rot}/N$ \\
\hline
GP & 13.334 & 9.889 & 0.3926 & 5.8056 & 3.445 & - &  0.2463 \\
     & (13.187) & (9.7836) & (0.4251) & (5.7427) & (3.4039) & - & (0.2119) \\
MGP & 15.811 & 11.443 & 0.3405 & 7.1248 & 2.4999 & 1.2454 &  0.2324 \\
     & (15.623) & (11.305) & (0.37692) & (7.0377) & (2.4824) & (1.2233) & (0.1849) \\
\hline
\end{tabular}
\end{center}
\end{table}

\begin{table}
\caption{Results for the chemical potential and the total energy per unit number of bosons  in units of $\hbar\omega_{\bot}^{0}$ obtained from the GP and the MGP calculations for the condensate of of $N = 10^{4}$ $^{85}$Rb atoms trapped in an anisotropic trap
with $\lambda_{0} = 0.39$ and $\omega_{\bot}^{0}= 2\pi\times 17.5Hz$. }
\begin{center}
\begin{tabular}{|c|c|c|c|c|}\hline
$a/a_{0}$ & \multicolumn{2}{c|}{GP} & \multicolumn{2}{c|}{MGP}  \\
\cline{2-5}
& $\mu_{1}$ & $E_{1}/N$  & $\mu_{1}$ & $E_{1}/N$  \\
\hline
1400 & 10.223 & 7.545 & 10.621 & 7.790  \\
3000 & 13.649& 9.955&  14.898 & 10.739  \\
8000 & 19.955& 14.422 & 24.798 & 17.528  \\
10000 & 21.774 & 15.715 &  28.223 & 28.223 \\
\hline
\end{tabular}
\end{center}
\end{table}
\clearpage
\newpage
\begin{figure}
\begin{center}
\end{center}
\caption{Comparison of the GP (dashed line) and the MGP (solid line) density profiles  of a vortex state with $\kappa = 1$ as a function of the radial distance $r_{\bot}$ (in units of $a_{ho}$)  from the z-axis for  $N= 10^{4}$ $^{85}$Rb atoms trapped in an anisotropic trap with $\lambda_{0} = 0.39$ and $\omega_{\bot}^{0}= 2\pi\times 17.5Hz$ . The scattering length $a/a_{0} = 1400$.}
\end{figure}
\begin{figure}
\begin{center}
\end{center}
\caption{Same as Figure 1 but with the scattering length $a/a_{0} = 10000$.}
\end{figure}
\begin{figure}
\begin{center}
\end{center}
\caption{The peak density $n_{peak}$ as a function of scattering length $a/a_{0}$ of a condensate with a quantized vortex of quantum of circulation $\kappa = 1$. The solid line corresponds to the MGP result and the dashed line is obtained with GP calculation. The trap parameters are same as Fig. 1.}
\end{figure}
\begin{figure}
\begin{center}
\end{center}
\caption{The peak position $r_{1{\bot}}^{peak}$ (in unit of $a_{ho}$) as a function of scattering length $a/a_{0}$ of a condensate with a quantized vortex of quantum of circulation $\kappa = 1$. The solid line corresponds to the MGP result and the dashed line is obtained with GP calculation. The trap parameters are same as Fig. 1.}
\end{figure}
\begin{figure}
\begin{center}
\end{center}
\caption{The frequency shift (in unit of $\omega_{\bot}$) of the  quadrupole oscillations of a BEC due to the presence of a single quantized vortex ($\kappa = 1$) as a function of scattering length $a/a_{0}$.  The solid line corresponds to the MGP result and the dashed line is obtained with GP calculation. The trap parameters are same as Fig. 1.}
\end{figure}
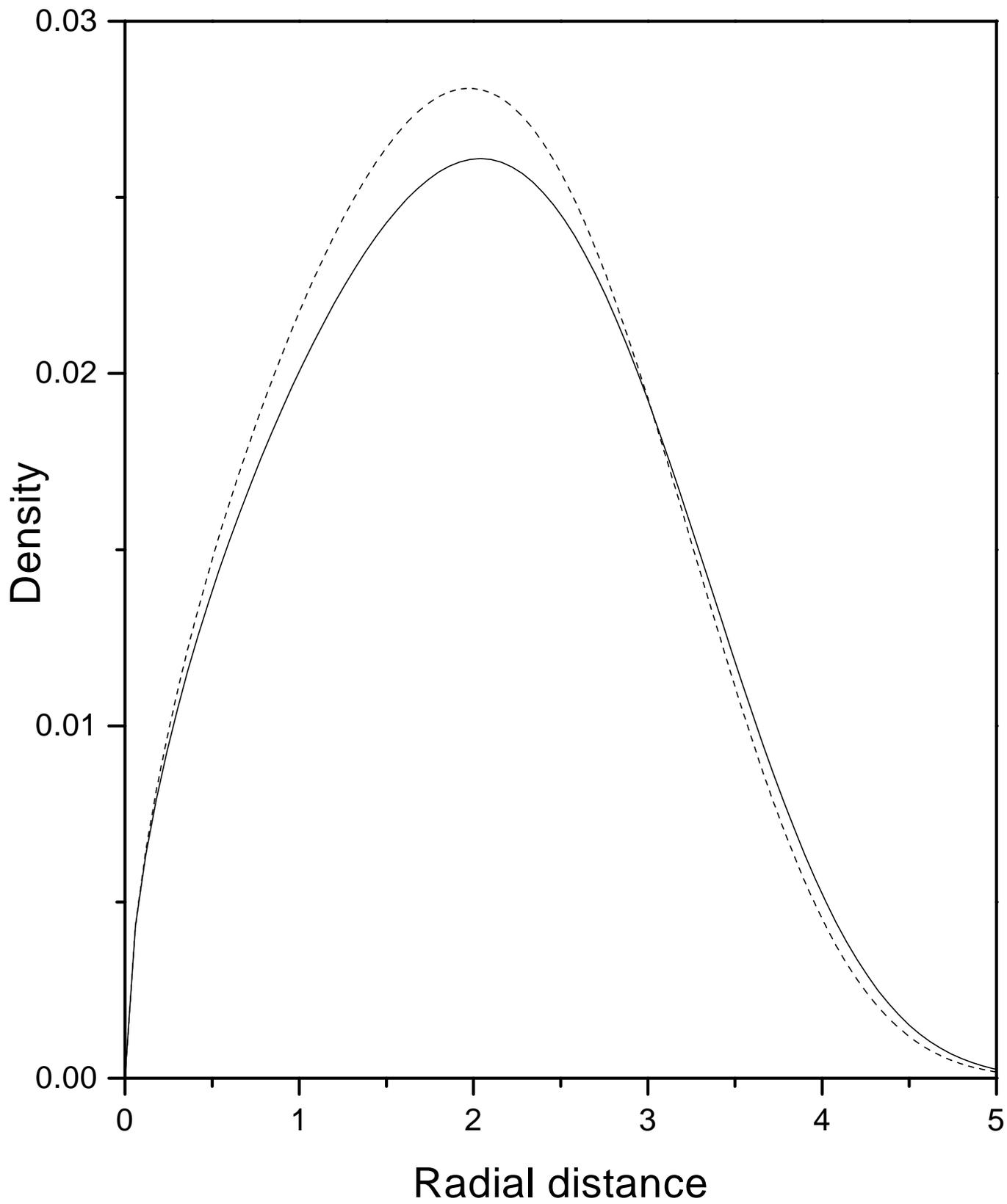
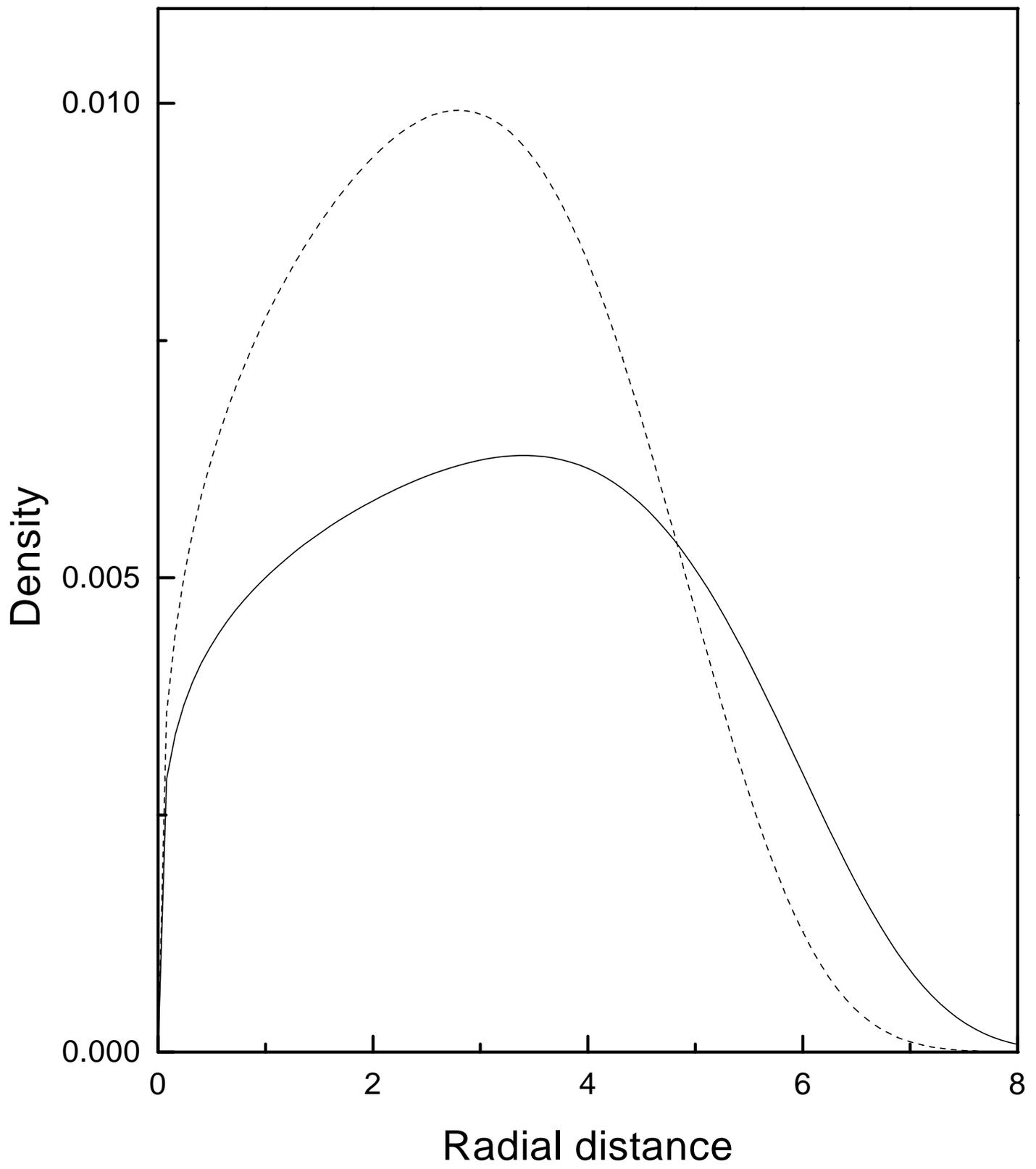
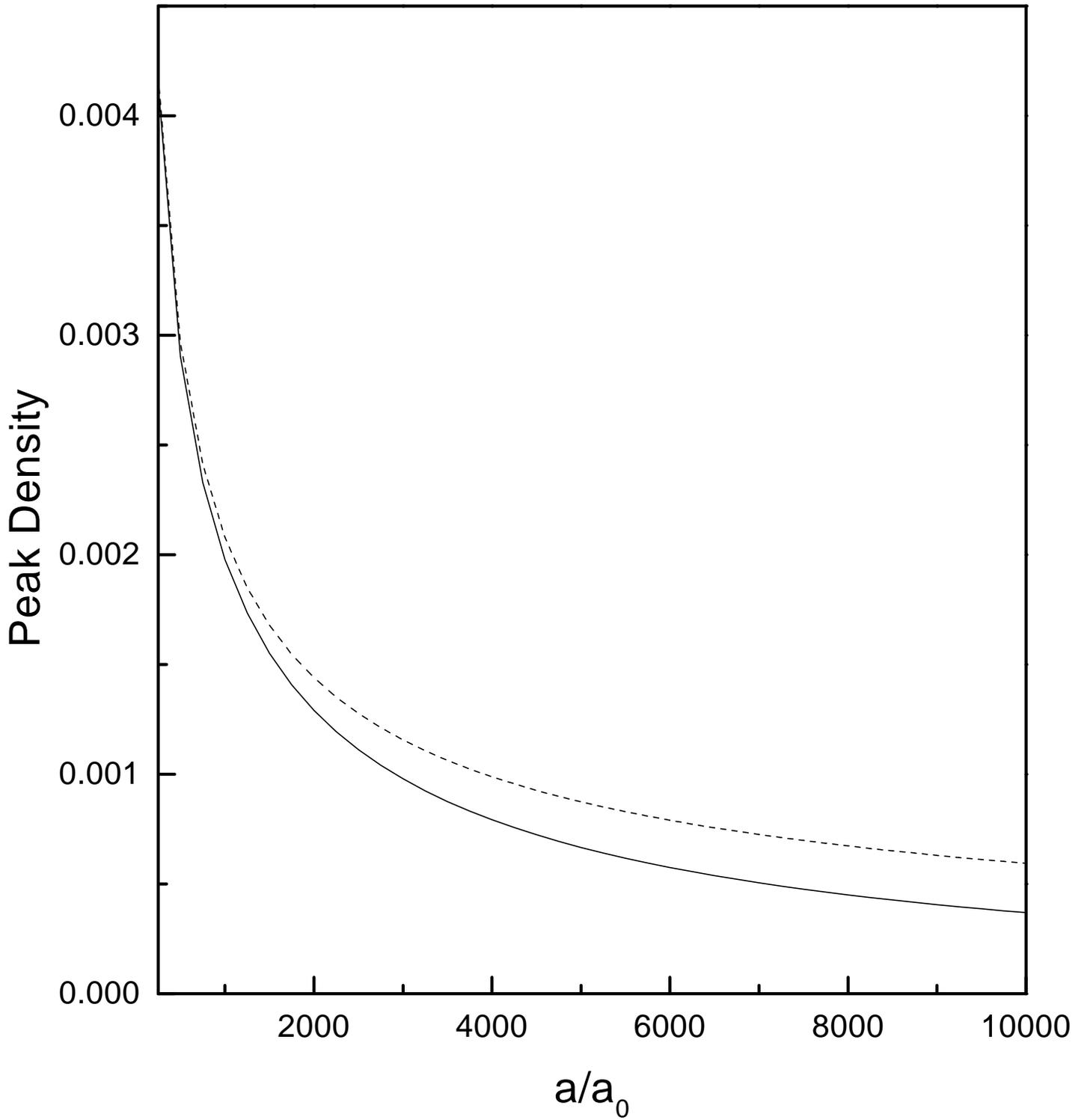
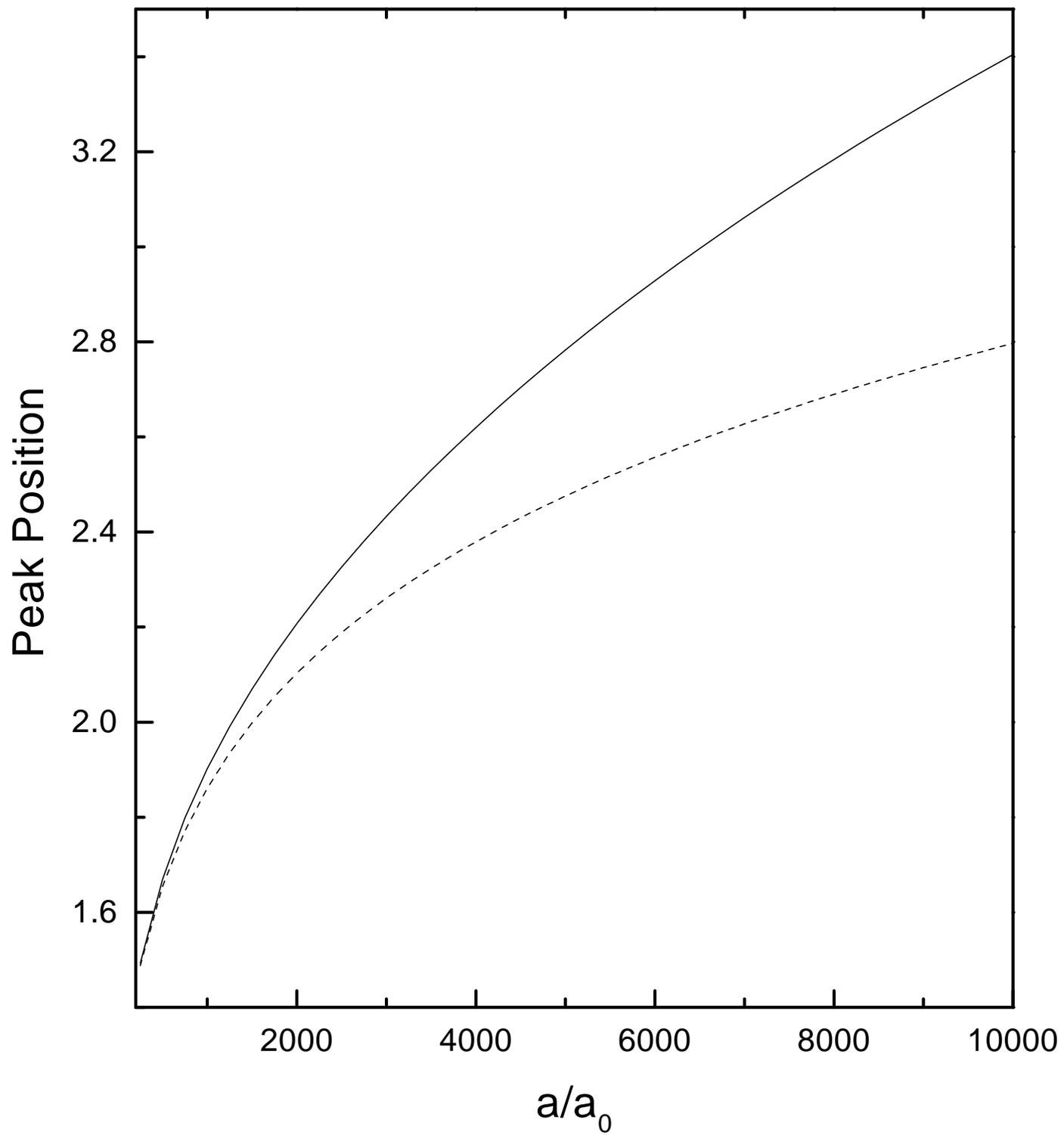
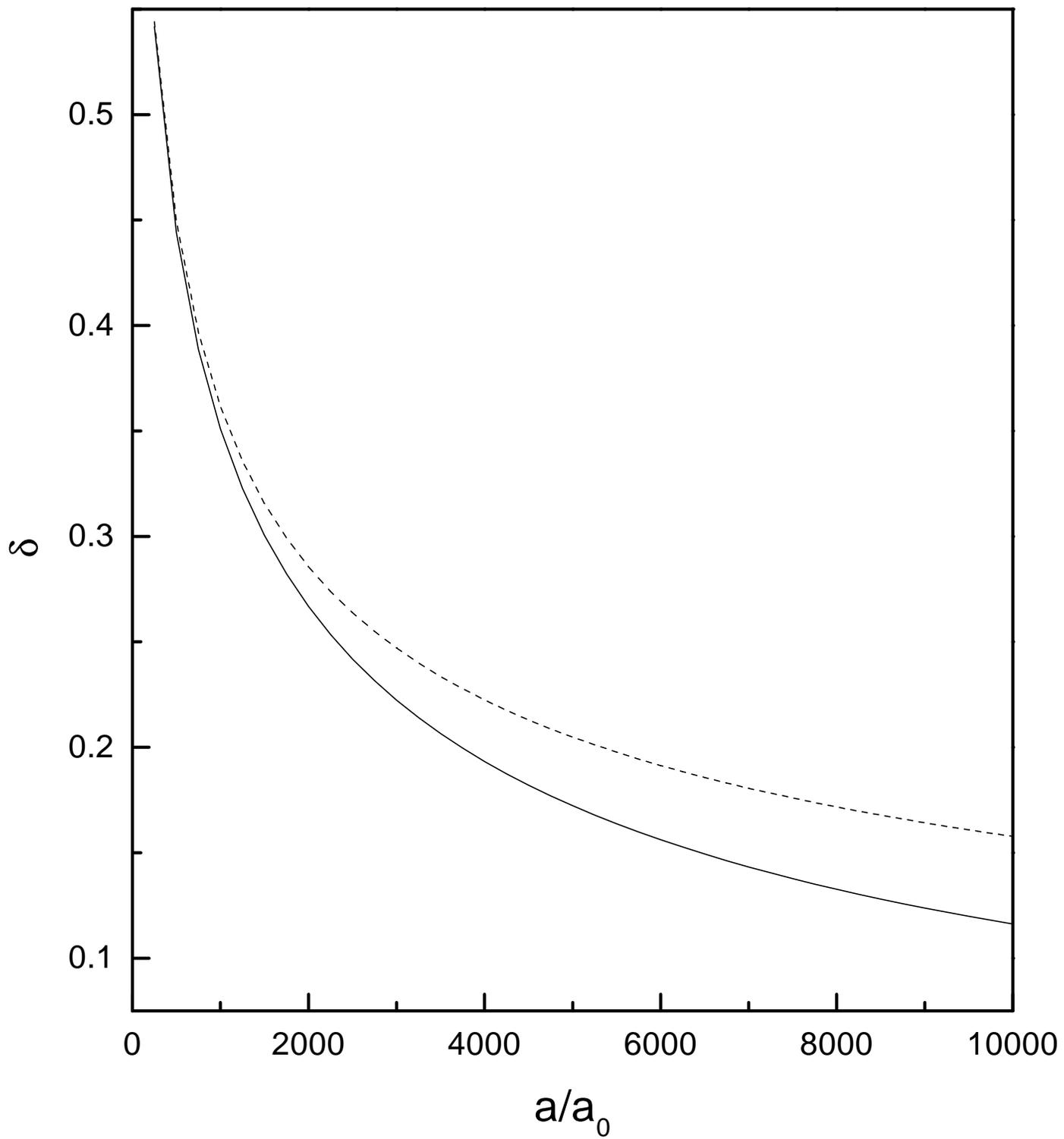
\end{document}